# An Efficient Approach for Calculating Free Energy in Molecular Dynamics: Demineralization of Hydroxyapatite as a Case Study

Mahdi Tavakol[1], Jin-Chong Tan[1*], and Alexander M. Korsunsky[2**]

[1] Multifunctional Materials & Composites (MMC) Laboratory, Department of Engineering Science, University of Oxford, Parks Road, OX1 3PJ, UK

[2] Professor and Fellow Emeritus, Trinity College, University of Oxford, OX1 3BH, UK

Corresponding authors:

[*]jin-chong.tan@eng.ox.ac.uk

[**]alexander.korsunsky@gmail.com

**Abstract**

Despite the unique strength of Molecular Dynamics (MD) simulations in providing novel insights into the microscopic details of phenomenon of interest in many fields in materials science, physics and biology, the biggest barrier towards its application is its limited timescale which is several orders of magnitude lower than the timescale of the real-world processes and phenomena being modeled. Free energy calculations are designed as a remedy to this problem that in theory can overcome this barrier. This is particularly relevant for biomineralisation processes such as tooth mineral formation and dissolution, while also reflecting a broader challenge in accurately modelling rare events and long-timescale phenomena across complex molecular systems. However, due to the novelty of the field, a number of questions remain outstanding pertaining to the best practice of applying this method. The non-equilibrium work approach based on the Jarzynski equation is one of the most promising free energy calculation methods. However, the biggest challenge for the widespread use of this method is the question of how many simulations are required for an accurate free energy estimation. Comparing the free energy results with very long reversible pulling simulations of atom clusters from the surface, each taking 75 days on the 48-core Intel Xeon Platinum 8268 CPUs, in this study we showed that this question is irrelevant and higher quality and better equilibrated initial structures is the proper approach than simply based on the number of simulations. Based on this significant observation, we designed a new adaptive free energy calculation methodology which combines high quality, high computational cost free energy values with lower quality, lower cost values to build up the entire free energy profile. In the best case scenario this method lowers manifold the computational cost required for the non-equilibrium work free energy calculations compared to both the regular method and the reversible simulation, while in the worst case it acts merely as a parallelized version of the reversible pulling method with the same computational cost. The presented method always reduces the sampling error in free energy calculations. In general, the present method paves the way for the application of free energy calculation in new areas of interest thanks to the combination of lower computational cost with increased accuracy of the method.

## 1. Introduction

Since the pioneering work of Alder and Wainwright on the simulation of liquid argon [1], molecular dynamics (MD) simulation has advanced over the last decade to the point where it has become an indispensable tool in physical chemistry, biophysics, and materials science. MD provides atomistic insights that complement experimental techniques and often reveals details inaccessible by direct measurement. Despite its success, conventional MD simulations are constrained by the timescales accessible to routine computation, which typically extend only to the microsecond regime at best [2]. Many processes of fundamental importance in chemistry and biology occur on much longer timescales, separated by high free energy barriers that are rarely crossed during conventional simulations in which no bias is applied to the system under study. Examples include ligand binding and unbinding in drug discovery [3], ion permeation through membrane channels [4], protein unfolding [5], nucleation phenomena in condensed matter systems [6], and amyloid misfolding [7]. Because such rare events underlie critical molecular phenomena, the ability to compute accurate free energy differences and profiles is of vital importance.

At the conceptual level, the importance of free energy stems from the fact that it governs the stability of states, determines equilibrium populations, and sets the likelihood of transitions. Unlike potential energy, which reflects only a single configuration, free energy integrates both energetic and entropic contributions, thereby accounting for accessible configurations. As such, free energy landscapes provide the most meaningful framework for describing molecular processes. A wide range of computational approaches have been developed to estimate free energies. Classical methods such as free energy perturbation (FEP) and thermodynamic integration (TI) rely on equilibrium sampling along either physical or alchemical pathways [3, 8] in which the system is gradually transformed between thermodynamic states through a parameterized change in the Hamiltonian rather than a physically realizable pathway. Umbrella sampling [9] and related weighted histogram analysis methods (WHAM) [10] improve efficiency by restraining simulations in overlapping windows across the reaction coordinate, allowing systematic reconstruction of the free energy surface. Enhanced sampling techniques such as metadynamics [11] and adaptive biasing force (ABF) [12] employ history-dependent or adaptive potentials to discourage resampling of previously visited states and thereby accelerate barrier crossing. More recently, nonequilibrium work relations, including the Jarzynski equality [13] and Crooks fluctuation theorem [14], have introduced a radically different perspective by linking distributions of non-equilibrium work to equilibrium free energy differences. Each of these approaches embodies distinct trade-offs in efficiency, accuracy, and applicability, but all share the common aim of extracting rigorous thermodynamic information from microscopic trajectories.

A critical issue in free energy calculations is the estimation of the errors arising from both the statistical and sampling errors. Since canonical MD simulations sample the energy according to the Boltzmann distribution, the convergence for various parameters of interest is implied by the convergence of their statistical error of mean. In other words, enough sampling of peak of the distribution surveyed by the statistical error (standard deviation or error of mean) implies the convergence for various parameters in conventional MD simulations, which is not the case for the free energy calculations for which the sampling error dominates. Sampling errors reflect incomplete sampling of the important regions of the phase space, particularly when states are separated by large barriers or when overlap between reference and target ensembles is insufficient. In umbrella sampling it manifests itself in poor overlap between neighboring windows, while in the non-equilibrium work approach of Jarzynski equality is due to the absence of rare-event works which has the largest weight in the ensemble average. Thus, this type of error is hard to estimate, and the careful evaluation of errors is necessary for the assurance of reliability of the calculated free energy values.

The Jarzynski equality introduced in 1997 relates the non-equilibrium work distribution with the free energy difference [13]. In practice, it elegantly implies that reversible work (e.g. free energy) can be calculated through a weighted average of non-equilibrium works. Even though accurate in theory, this method suffers from serious problems in sampling important events. Since the equality involves the ensemble average of exponential irreversible works it effectively gives higher weights to the lower work values which themselves are kind of rare events. More specifically, the MD simulation samples the work distribution while the Jarzynski equation samples the exponential of work and the tail of the former distribution coincides with the peak of the later [15]. The overlap between work distribution and the free energy estimator affects the quality of free energy predictions. Thus, estimators such as the Bennet Acceptance Ratio (BAR) and the second-order approximation [16] have been developed to increase the accuracy through approximating the formula with the ensemble average of other parameters with peaks closer to the work distribution. The BAR method combines the forward and reverse work distribution while the second-order approximation deploys the first and second moment of the distribution [16], both of which improve the overlap between these two distributions. Since the non-equilibrium work is sampled from initial equilibrium distribution, the accuracy of the free energy is dependent on both the size and sampling accuracy of this initial equilibration. Thus, the questions about the application of Jarzynski are what the size of this initial equilibration should be, how should it be sampled, and finally how the free energy should be calculated from it, all of which formed the subject of the current study.

In summary, there are several questions on how to apply the elegant Jarzynski equation which is investigated in the current study for the biomineral Hydroxyapatite (HAp) as the case study. The results presented in the current study show that despite the common belief, running more simulation does not always translate to more accurate free energy values while how the initial structures for non-equilibrium work are created matters more. To reflect this observation, a new free energy method designated as adaptive-BAR is developed, which reduces the error significantly. The application of this method for free energy calculations in various contexts can lead to great breakthroughs due to its lower computational cost.

## 2. Computational methods

In the current study the Hydroxyapatite (HAp) mineral surface was used as the case study to investigate the different non-equilibrium pulling approaches in calculation of the free energy. In the first section the system setup for MD simulations of the HAp is described. Then, in the second section the non-equilibrium approach used in the current study is summarized including how the initial structures were created and how the free energy calculations were done.

### 2.1. System setup

The system under study which is composed of HAp is shown in Figure 1a. The initial HAp structure was taken from the Interface forcefield database [17] and it was solvated in the TIP3P water model. The Interface forcefield was also utilized for HAp interactions with water molecules. HAp facets of (010) and (001) were considered under pH values of 5 and 7. Various elements of HAp including ($Ca^{2+}$(I), $Ca^{2+}$(II), $PO_4^{3-}$, $OH^-$, $Ca^{2+}$(surface) and $PO_4^{3-}$ (surface) were selected for being pulled in non-equilibrium MD simulations. More information on the system setup is available in our previous study [18].

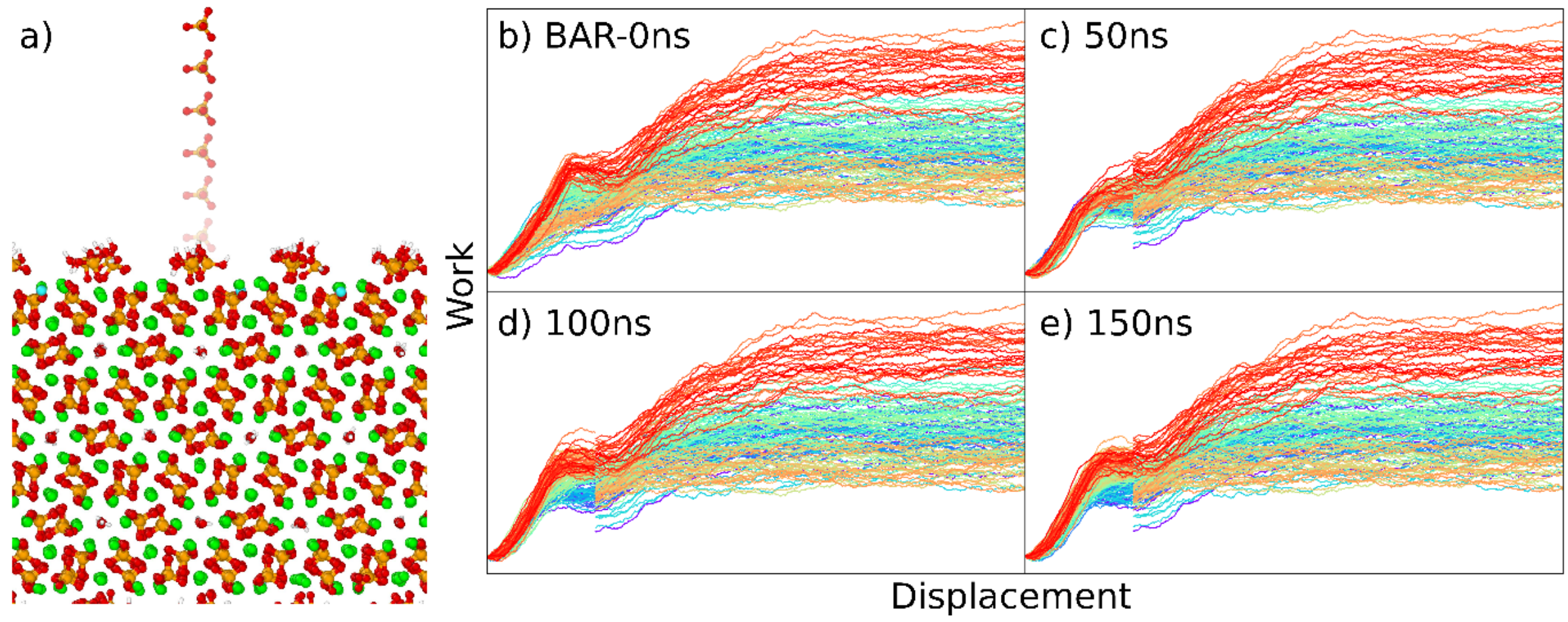

*Figure 1 – The molecular dynamics (MD) simulation setup. a) The HAp under study with one of the pulled atoms of surface phosphate ions shown. The work distribution for b) the common free energy approach in literature and the suggested adaptive free energy calculation method with c) 50 ns, d) 100 ns and e) 150 ns equilibration length before creating the initial frames for the non-equilibrium atoms pulling simulations. Panels (b) – (e) are plotted on the normalized work and displacement axes.*

## 2.2. Free energy calculations

The Jarzynski method for the free energy calculations was deployed according to which there is an equality relationship (equation 1) between the non-equilibrium work and the free energy for which the work values were calculated through equation 2. Even theoretically correct, this equality suffers from very slow convergence limiting its practical applications. Thus, two other methods of Bennet Acceptance Ratio (BAR, equation 3) and the second-order approximation (equation 4) have been suggested in the literature. In these equations the $\Delta G$, $W_F$, $W_R$, $\langle \quad \rangle_F$ and $\langle \quad \rangle_R$ terms represent respectively the free energy, forward work, reverse work and ensemble averages of works obtained from the structures sampled from the equilibrium simulation started from the beginning and the end of the pulling simulation. First, these three estimators were compared in the current study.

$$e^{-\frac{\Delta G}{kT}} = \langle e^{-\frac{W_F}{kT}} \rangle_F \tag{1}$$

$$W_F = \int_F f dx \approx \sum_{i=0}^{t} \left( \frac{f_i + f_{i+1}}{2} \right) * (x_{i+1} - x_i) \tag{2}$$

$$e^{-\Delta G/kT} = \frac{\left\langle \frac{1}{1 + e^{W_F - \Delta G}} \right\rangle_F}{\left\langle \frac{1}{1 + e^{W_R + \Delta G}} \right\rangle_R} \tag{3}$$

$$\Delta G = \langle W \rangle_F - \langle (W - \langle W \rangle_F)^2 \rangle_F \quad (4)$$

As the free energy calculation involves an ensemble average of a work-related parameter, there is a need for a proper sampling of initial structures to start irreversible pulling from. To this end, the equilibrium simulations were started with different initial random seed numbers. At least three different random number seeds were used, and for each the system was equilibrated for at least 30 ns or 50 ns, and the simulations configurations (particle positions + velocities) were outputted every 1 ns resulting in at least 3x30 simulations. For each of the configurations one pulling simulation was started to obtain the ensemble average.

To obtain a basis to compare the free energy values with and calculate the errors in the free energy values, for each pulling case (different combinations of pulled ions, facets and pH values) at least one very long pulling simulations with pulling rate of 0.01 Å/ns and at least three long pulling simulations under a pulling rate of 0.1 Å/ns were done. Each of the long reversible simulations took around 80 days on the 48 core Intel Xeon Platinum 8268 CPUs.

Comparing these reversible works with the free energy values obtained from these three estimators, a huge difference was observed. To improve the quality of the free energy estimates, two different approaches of increasing the number of simulations and increasing the equilibration length were tested. In the first case, at least 10 initial random seed numbers were chosen for 10 different equilibration simulations, each lasting for 30 ns leading to at least 300 pulling simulations.

However, for the second case based upon which the adaptive free energy calculation was built, instead of using frames 0-50 ns (or 0-30 ns) for the pulling simulations the first *n* frames were discarded, extracting the same number of frames though increasing the length of the equilibrium simulations. In the adaptive free energy calculations, the results of the BAR and the shorter pulling simulations with longer equilibration simulations were mixed which is explained later on (Figure 1c-e).

## 3. Results

The free energy values for pulling each one of the $Ca^{2+}$(I), $Ca^{2+}$(II), $PO_4^{3-}$, $OH^-$, $Ca^{2+}$(surface) and $PO_4^{3-}$ (surface) ions from the HAp (001), (010) at pH values 5 and 7 and calculated with three different estimators are shown in Figures 2, S1-S4. The results for the second-degree approximation are not shown in these figures since the errors are exceedingly high. Comparing the exponential formula results with BAR, the average of forward-reverse works and the reversible works, it is deducted that the BAR method leads to more accurate results than the exponential estimator. For instance, in the case of Ca(I) in figure 2a for the distances higher than 3 Å were the forward and reverse average works equal to each other and the reversible work accordingly, the exponential formula shows large deviations from this value despite the BAR output. The same observation can be made for pulling the $Ca^{2+}$(surface) and $PO_4^{3-}$(surface) as shown in figure 2.

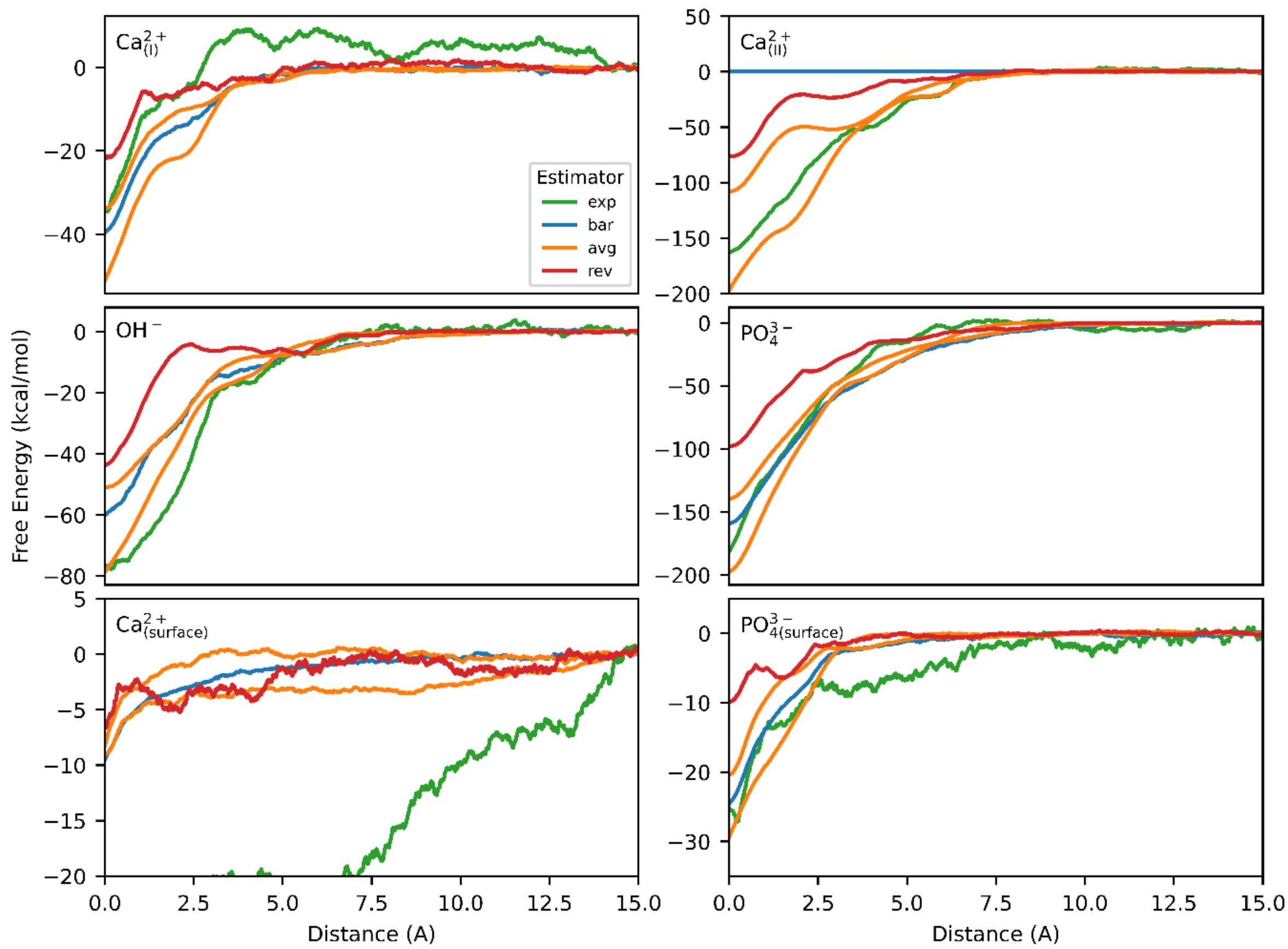

*Figure 2 – Comparison of different free energy estimators with the reversible work for pulling different ions from the (001) facet of HAp at pH5.*

The effect of varying the number of pulling simulations for cases of pulling $Ca^{2+}$(surface) and $PO_4^{3-}$(surface) ions from (001) facets of HAp at pH values of 5 and 7 are presented in figure 3. These cases were chosen since lower free energy values of these cases makes the results easier to converge. To eliminate the effect of the random work distribution, 50 free energy values are reported for X number of simulation which is calculated from taking X work value out of whole simulations done. As the number of choices of taking X works values are prohibitively large, only 50 free energy values for each number of simulations. The results for pulling $PO_4^{3-}$ (surface) from the (001) facet HAp at pH5 (figure 3a) and pH7 (figure 3b), and $Ca^{2+}$(surface) from the (010) facet at pH5 (figure 3c) and pH7 (figure 3d) depicts that running more simulation will not result in reducing the difference with the reversible pulling free energy values (dashed lines in figure 3). This motivates us to present a new method to run non-equilibrium free energy calculations which is explained below.

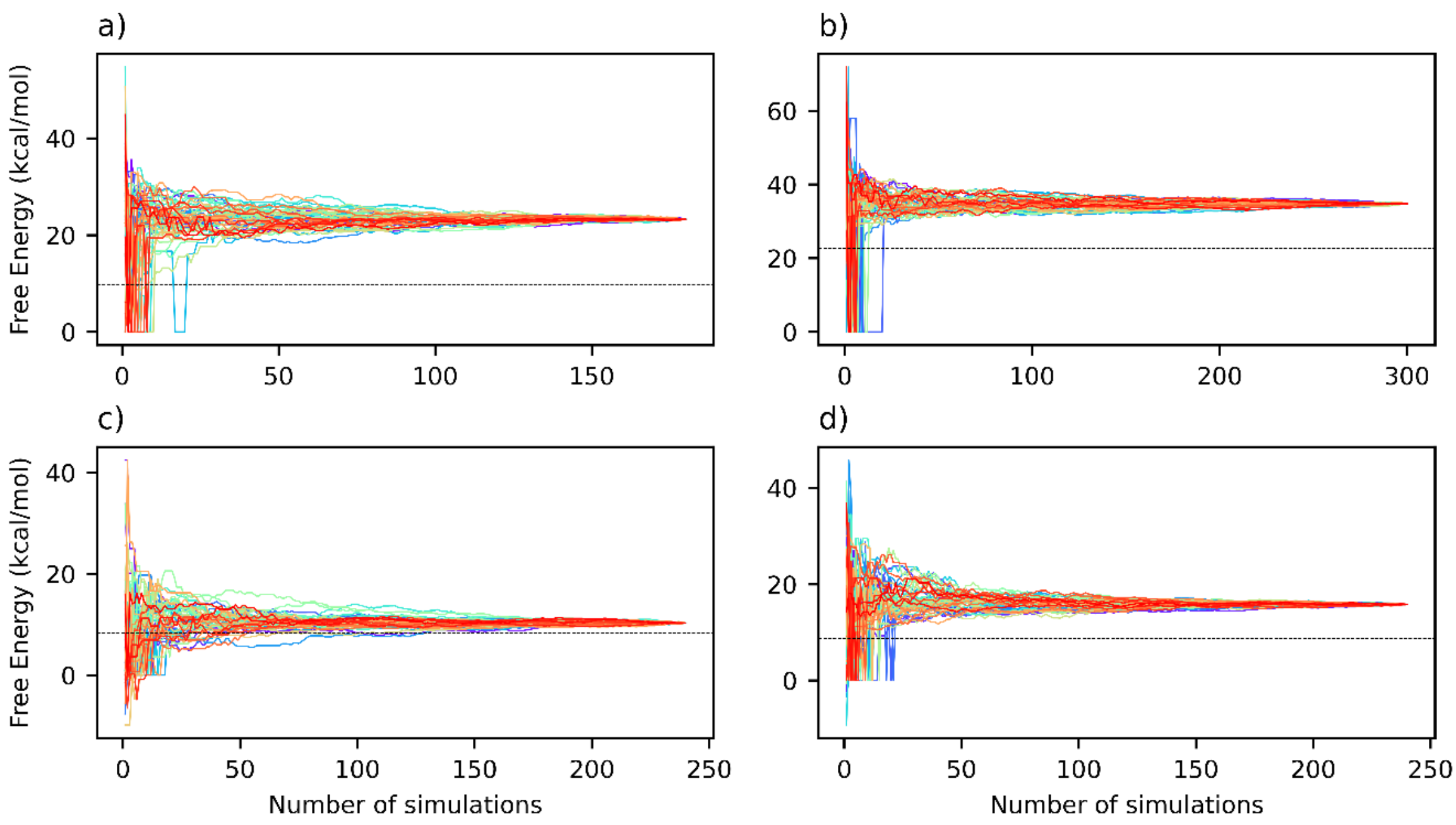

*Figure 3 – The effect of the number of simulations on the free energy values. The cases of $PO_4^{3-}$(surface) from the (001) surface at (a) pH5 and (b) pH7, and $Ca^{2+}$(surface) at (c) pH5 and (d) pH7.*

The idea behind the new method is motivated by comparing the simulation results for pulling $Ca^{2+}$ from the HAp (010) at pH5 (figure 4). Not surprisingly, the free energy value calculated from the BAR method has a large error of 67%. As the work values used in the free energy calculations are accumulated in the pulling pathway according to equation 1, an error in specific region will be accumulated and transferred to subsequent points in the pulling pathway. Besides, the free energy is a potential obtained through integration summed with an unknown integration constant and thus the free energy plot can be shifted upwards or downwards. To obtain the critical region, first we move the free energy plot to align its beginning with the reversible pulling (BAR in figure 4a) and another time we align the end points (BAR-shifted in figure 4b). For the BAR method, the beginning of the curve for a small region within 0-0.5 Å is in agreement with the reversible pulling results. While for the BAR-shifted the region after 2-3 Å is very close to the 0.01 Å/ns results, there is a small deviation in 3-5 Å range and after 5 Å its result is in a good agreement with the basis case. Thus, there are two critical regions of 0.5-2 Å and 2.5-5 Å for which the errors given by BAR method are high. Interestingly, in these regions the force obtained from average work or the BAR free energy values are high. Thus, we introduce a new non-equilibrium method called adaptive-BAR in which the force values are used to ascertain the critical region for free energy calculations and in this region either more simulation or simulations with longer equilibration stages are used to increase the quality of sampling. The choice of running more simulations versus longer equilibration and test cases for this method are discussed in the next paragraphs.

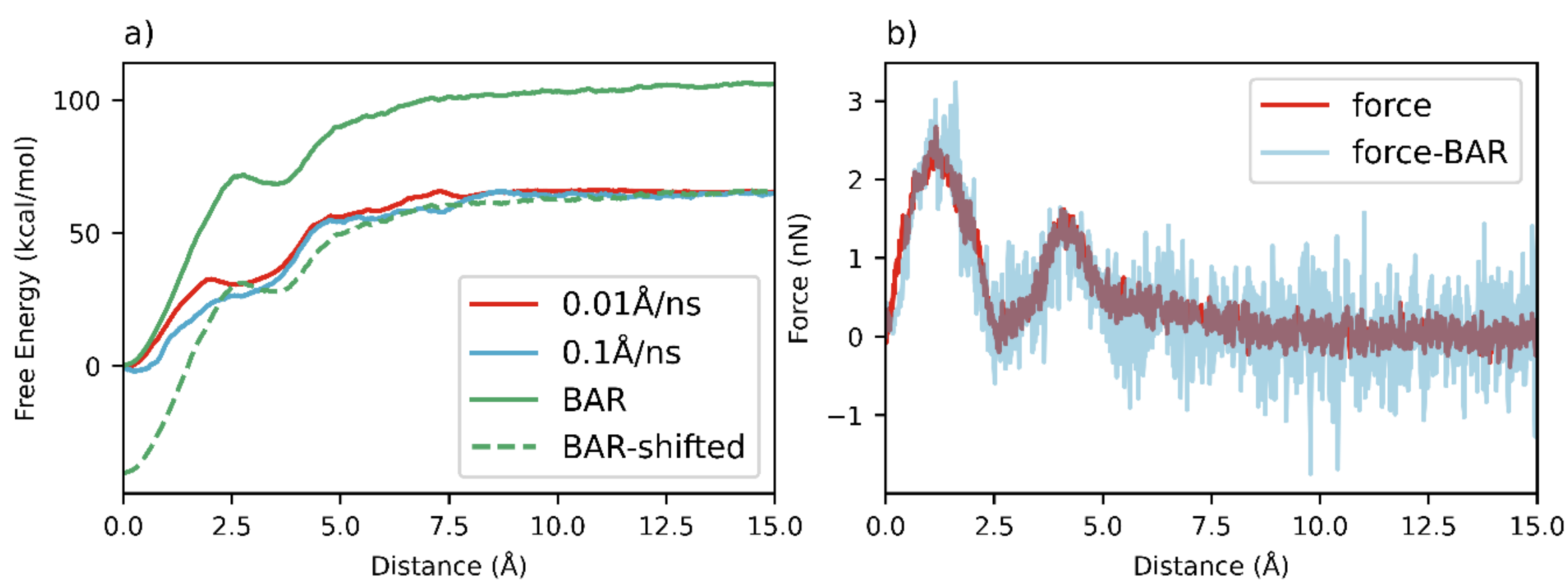

*Figure 4 – The comparison of the BAR results with reversible pulling free energy values. (a) The free energy profile depicts a distance in which neither the BAR nor its shifted curve correspond to the reversible free energy values which interestingly aligns with the region in which (b) the force obtained from the BAR method is large (0-3 Å).*

Here the adaptive-BAR simulations with more simulations in the critical region of 0-3 Å is compared with the adaptive version with longer critical region equilibration length (figure S4). The result shows that both running more simulations or having simulations with longer equilibrium improve the quality of the results. This result points to the idea that running Steered Molecular Dynamics (SMD) simulations with longer equilibration length in the critical region reduces the error. But the question is how long the equilibration should be and will its result finally match the reversible pulling free energy value.

To answer these questions and identify the practical aspects of the new method, it is tested by pulling $PO_4^{3-}$ and $Ca^{2+}$(I) from the (010) facet, OH- and $PO_4^{3-}$(surface) from (001) facets at pH5 and $PO_4^{3-}$(surface) from (001) facet at pH5 to scrutinize pulling various ions either in the surface or buried deep inside from various facets at different pH values. The first case of $PO_4^{3-}$ (surface) from (001) at pH5 (figure 5a) proves the effectiveness of the method. In this figure the adaptive method results with various equilibration lengths are compared with the reversible pulling results. As the variation in the work for the reversible pulling results is high, we use the lowest value as the reversible work to be compared with the adaptive method. This result shows that increasing the equilibration length reduces the difference between the adaptive bar results with the reversible work. For the case of pulling $PO_4^{3-}$ from (001) facet at the same pH level (figure 5b) despite having larger free energy difference the adaptive bar method performs better than the previous case. In this case the critical region is 0-6 Å and considering the reduced critical region of 0-3 Å will not lead to any improvement in the adaptive-BAR results. Other cases of $Ca^{2+}$(I)-010-pH5, $OH^-$-001-pH5 and $PO_4^{3-}$(surface)-001-pH7 also show notable improvements. For the last two cases the free energy values are lower than our reversible pulling results which shows less irreversibility and better results accordingly. The comparison of all cases with the reversible pulling generally shows a deviation from the reversible work (Figure 5c) and very large errors for BAR (Figure 5d) for most of the cases considered here. However, for the adaptive-BAR the error value decreases with increasing the equilibration length (figure 5d) before it is reaching a plateau. In a couple of cases the error for the adaptive-BAR reached the -10% to 10% region from the initial 50%-150% and there are cases that the adaptive-BAR outperforms the reversible pulling results reaching negative errors implying less reversibility.

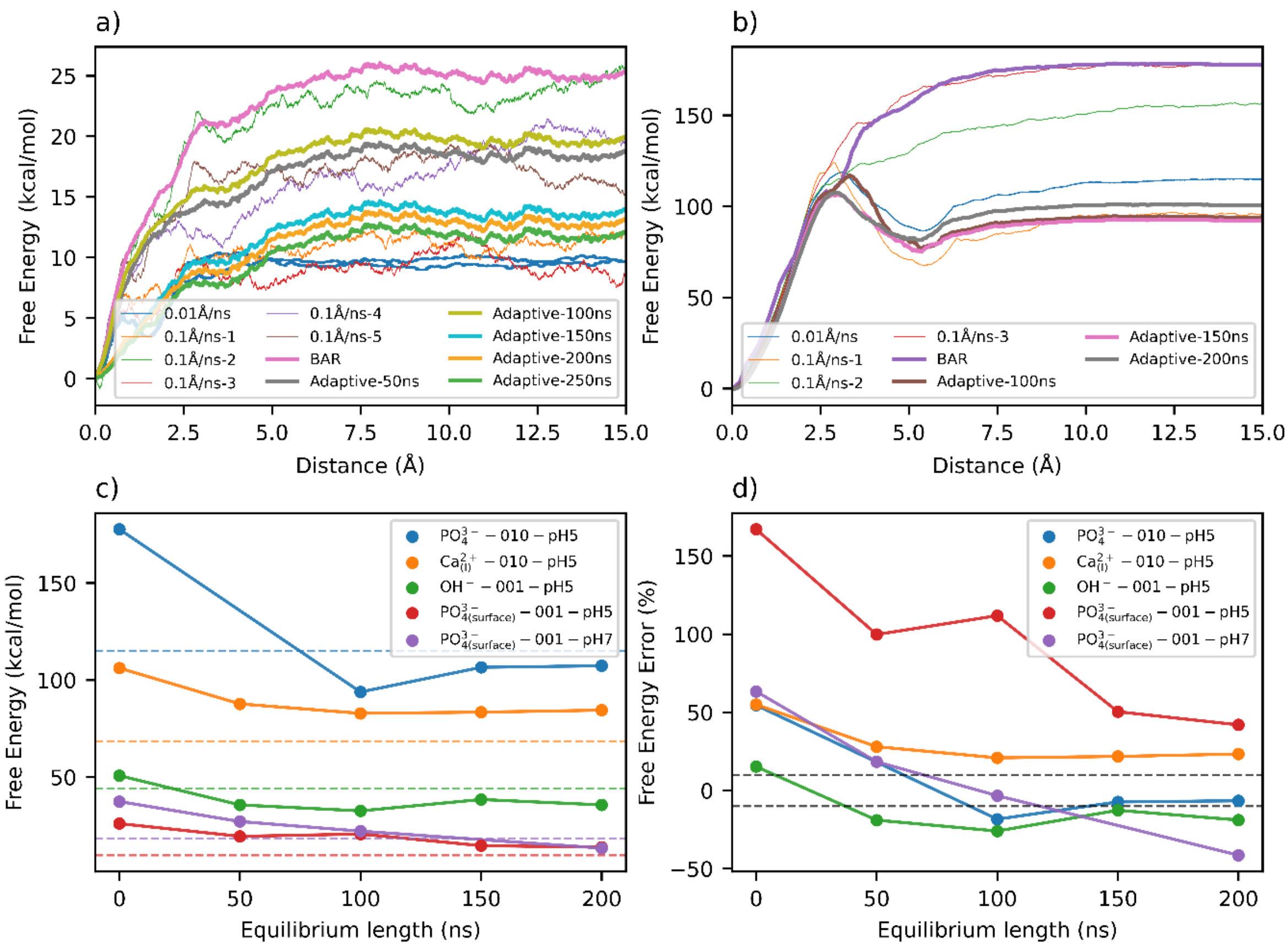

*Figure 5 – Comparison of the free energy calculations from BAR and adaptive-BAR with various equilibration length and reversible pulling results. The results for pulling (a) $PO_4^{3-}$(surface) from the (001) facet and (b) $PO_4^{3-}$ from the (010) facet of HAp at pH5. The variations of (c) free energy value and (d) error values with the length of equilibration run of adaptive-BAR for various cases tested in the current study. The equilibration length of zero corresponds to the traditional BAR method.*

Even though it is observed that juxtaposition of shorter MD simulations with longer equilibrium length with the BAR calculations increases the free energy calculation quality, the increased equilibration length adds to the computational cost. The computational cost of the adaptive method compared with the reversible pulling (0.01Å/ns) is shown in table 1 with the hypothesis that 3 reversible pulling are done and 150 non-equilibrium pulling are used for the free energy calculations. The computational cost of the adaptive-BAR though higher than the original BAR method, is still relatively lower than the reversible pulling result (Table 1).

*Table 1 – The computational cost of the adaptive method compared to the reversible pulling case. The cost/reversible ratio compares the BAR method computational cost to the cost of reversible pulling simulation.*

| Simulation | Equilibrium cost | Pulling cost | Critical cost | Total cost | Cost/Reversible |
|---|---|---|---|---|---|
| BAR | 3*50ns | 150*30ns | | 4650 | 0.517 |

| Adaptive-50ns-(0-3 Å) | 3*50ns + 3*50ns | 150*30ns | 150*6ns | 4800 | 0.533 |
|---|---|---|---|---|---|
| Adaptive-100ns-(0-3 Å) | 3*100ns+3*50ns | 150*30ns | 150*6ns | 4950 | 0.550 |
| Adaptive-150ns-(0-3 Å) | 3*150ns+3*50ns | 150*30ns | 150*6ns | 5100 | 0.567 |
| Adaptive-200ns-(0-3 Å) | 3*200ns+3*50ns | 150*30ns | 150*6ns | 5250 | 0.583 |
| Adaptive-250ns-(0-3 Å) | 3*250ns+3*50ns | 150*30ns | 150*6ns | 5400 | 0.60 |
| Reversible | | 3*3000ns | | 9000 | |

## 4. Discussion

Free energy calculations is a central concept in thermodynamics deployed to predict the spontaneity of a process. In molecular and material science, free energy calculation provides its practitioners with the ability to understand the chemical stability, binding affinity and reaction mechanisms at the molecular scale [19, 20]. These insights have a vast range of practical applications in drug discovery, biomaterials design, renewable energy storage and the development of efficient catalysis. Understanding free energy landscapes is essential for predicting how molecular systems evolve over time, particularly in complex biological and material environments. However, accurately estimating free energy differences remains computationally demanding due to the need for extensive sampling and the presence of metastable states. Therefore, improving the efficiency and reliability of free energy estimators is a key challenge across many fields of science and engineering. Among the various approaches employed in free energy calculations, the non-equilibrium work approach is based on the Jarzynski equation which relates the free energy, e.g. reversible work, with the non-equilibrium works [20, 21]. The equation, even though accurate in theory, has limited practical application due to difficulties in sampling works. Accordingly, several estimators have been designed for better sampling of works. However, it is not clear which estimator is more appropriate and how many simulations are needed despite various application of this method in the literature [21, 22]. In the current study this topic was investigated. In the context of biomineralisation, such as tooth mineral formation and dissolution, free energy calculations provide critical insight into the stability of mineral phases and their interactions with surrounding biomolecules and ions. Processes such as nucleation, growth, and demineralisation of hydroxyapatite are governed by subtle free energy differences that are difficult to probe experimentally. Therefore, robust computational methods are essential for understanding and predicting these mechanisms at the molecular level.

Among the three different estimators tested in the current study, it was observed that in most of the cases considered here the BAR has the lowest error in the free energy difference and the least scatter in the free energy profile. However, the error values as large as 70% was observed with respect to the reversible pulling cases. Increasing the number of simulations from 150 to 450 did not solve the problem, and indeed no noticeable difference was achieved by this brute force approach. Instead, the preliminary results showed that increasing the equilibrium length is more effective in improving the accuracy. This observation motivated the development of the adaptive-BAR method.

The adaptive-BAR method is motivated by the observation of the highest errors in the regions of the pulling pathway with larger BAR forces called critical regions. Thus, in the new adaptive method, first the critical regions are identified and for each starting structure a longer equilibration is done, then from these equilibrated structures shorter pulling simulations are performed (the length of the critical region) and the original BAR is combined with these new results for the critical region. For the cases considered here a large reduction in the error values with respect to the reversible pulling is observed and for the most extreme cases the free energy goes below the slow pulling case (0.01Å/ns), ran for around 75 days, which is impressive showing less irreversibility than the very slow simulation. Even though the authors observed an improvement with increasing the equilibration time for the critical region for the case of HAp system, they cannot exclude such a possibility with running more simulations in that region for other systems. Running longer equilibration is effective since during reversible pulling there is an equilibration in the dimension perpendicular[1] to the pulling direction which longer equilibration for the critical region can incorporate into the free energy calculations. However, the nature and timescale of this relaxation process have not been previously characterised.. The long simulation time for the reversible pulling naturally incorporates these changes into the adaptive-BAR which is absent from non-equilibrium simulation with shorter equilibrium time. Two important points should be mentioned here: (a) the correct selection of the critical region matters very much as for the figure 5b by selecting 0-3 Å region instead of 0-6 Å does not provide any benefit for the adaptive method, (b) after a specific threshold increasing the equilibration length will not improve the adaptive method quality and thus it is deduced that the system is equilibrated well enough for the free energy calculations. It is worth noting that the definition of equilibration is dependent on the system and the property under study. The equilibration for dipole moment for example might take longer than its energy [23].

[1] Here by the perpendicular direction we meant the dimension in the 6*N* space where *N* is number of atoms rather than geometrical dimensions.

Thus, for the first time we (1) highlighted the importance of equilibration for non-equilibrium free energy calculation and (2) effectively, suggested a criterion for this purpose.
Thus, the current method suggests running a conventional BAR simulation to obtain the critical region and for critical regions another shorter bar simulations with longer equilibration length is run and then the equilibration length is increased until a plateau is determined in the free energy plot with respect to the equilibration length. It is true that running BAR with longer equilibration would have the same effect but finding the required equilibration length with the bar requires repeating the simulation for the whole profile rather than the critical region as in our method, consuming more CPU-hours. After finding the required equilibration lengths two different approaches can be followed: Either (1) mixing the results with the original BAR with the critical region BARs with longer equilibrium lengths or (2) running a final BAR simulation with those longer equilibration lengths. Here, we just tested the first approach, but the authors do not have any reason to believe that the latter does not work.
Even though this work sheds more lights onto unexplored aspects of free energy calculations, like any other studies it is not devoid of possible shortcomings. First, the present analysis is carried out for HAp as a case study. While the principles are general, the quantitative behavior may differ for proteins, membranes, or other complex systems. Anyway, this method will be tested on other systems in further studies. Another shortcoming is related to the reversible pulling results used to verify the free energy calculations. Even though the pulling rate here was 0.01 Å/ns and involved 75 days in each direction of forward and reverse pulling (150 days), the work values except for pulling surface atoms did not match for these direction so there is still irreversibility in these cases too. The authors hope that in the near future with the advancement of more powerful hardware this shortcoming will be solved. Finally for every system under periodic boundary conditions, there is a chance of periodic box effect for which the authors believe should be minimal for this system considering the fact that it is made of several HAp lattices.

## 5. Concluding remarks

In the present study, a new method for free energy calculation based on the Jarzynski equality was suggested. The simulation results illustrated that despite the common idea that increasing the number of simulations is the way forward towards more accurate free energy calculations, we showed that in contrary, the higher quality and better equilibrated initial structures for the non-equilibrium work sampling, is the proper approach. According to this observation a new free energy method was suggested which involves first running a low-cost free energy calculation with either low quality initial structures or faster pulling simulations to obtain the critical regions of the free energy profile which involves large changes in the free energy profile (high force value). Subsequently, for these regions either (1) higher quality initial structures or (2) more simulations are used for shorter free energy calculations. Then the results are merged and these stages are repeated until the free energy converged. The results illustrate that using higher quality initial structures obtained from longer equilibrium simulations is the effective scenario. For the other case, we did not see any evidence in favor of increasing the number of simulations. The new method reduces the sampling error and the computational cost of the original method which can widen the area of application of this method to new applications.

## 6. Data availability

The data required to generate the results presented in this paper are included in the inputData.zip file provided in the supplementary materials.

## 7. Acknowledgements

Via our membership of the UK's HEC Materials Chemistry Consortium (MCC) funded by EPSRC (EP/X035859), the present work used the ARCHER2 UK National Supercomputing Service (http://www.archer2.ac.uk). We also acknowledge the use of the University of Oxford Advanced Research Computing (ARC) facility for carrying out this work (https://doi.org/10.5281/zenodo.22558).

## 8. Funding

AMK and all the co-authors wish to acknowledge the UK Engineering and Physical Sciences Research Council for the award of the principal funding for the present study under EPSRC grant no. EP/W009412/1 “Elucidating the pathways for human tooth enamel mineralization by 4D microscopy and microfluidics”.

## *Supplementary Information*

# An Efficient Approach for Calculating Free Energy in Molecular Dynamics: Demineralization of Hydroxyapatite as a Case Study

Mahdi Tavakol[1], Jin-Chong Tan[1*], and Alexander M. Korsunsky[2**]

[1] Multifunctional Materials & Composites (MMC) Laboratory, Department of Engineering Science, University of Oxford, Parks Road, OX1 3PJ, UK

[2] Professor and Fellow Emeritus, Trinity College, University of Oxford, OX1 3BH, UK

Corresponding authors:

[*]jin-chong.tan@eng.ox.ac.uk

[**]alexander.korsunsky@gmail.com

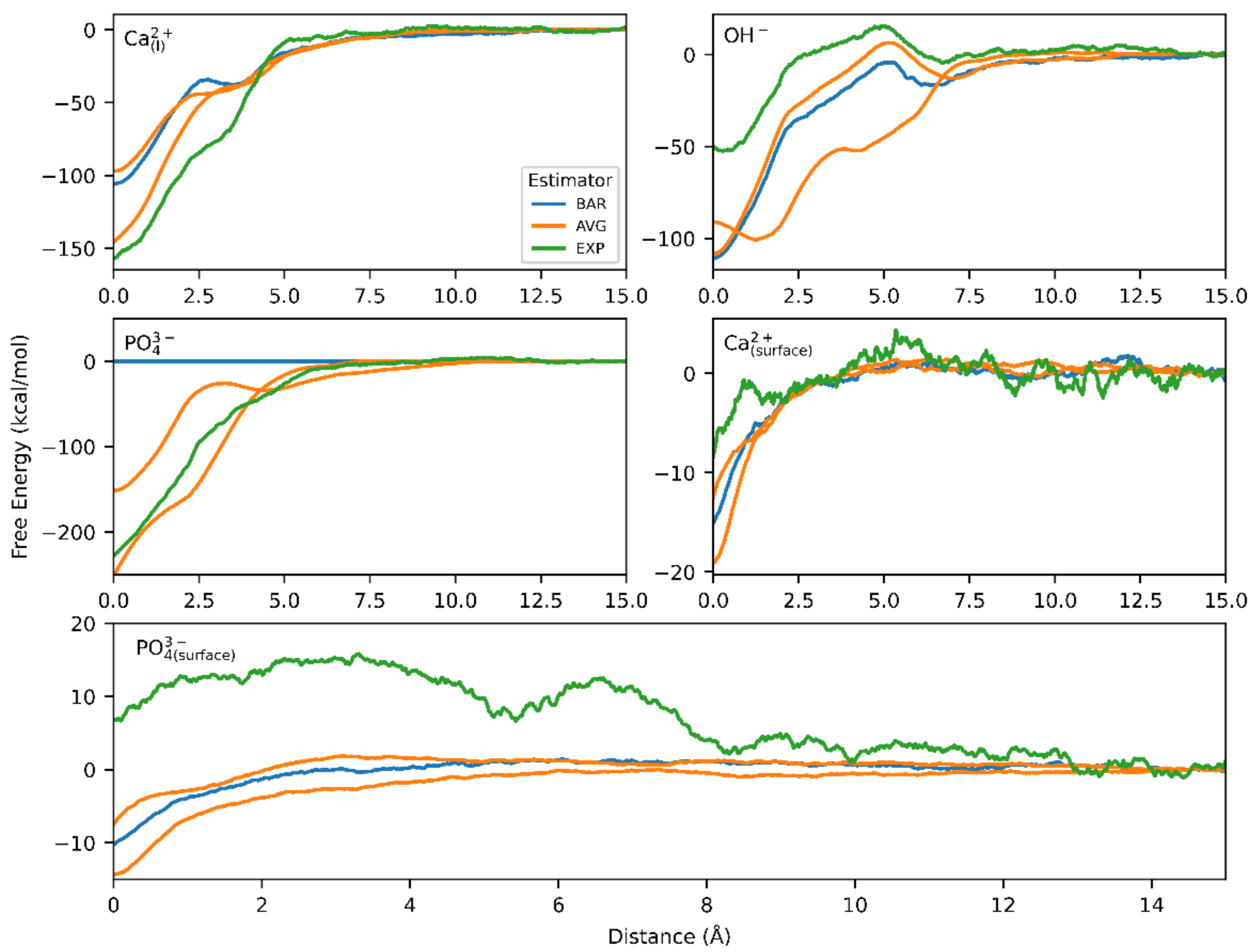

*Figure S 1 - Comparison of different free energy estimators with the reversible work for pulling different ions from the (010) facet of HAp at pH5.*

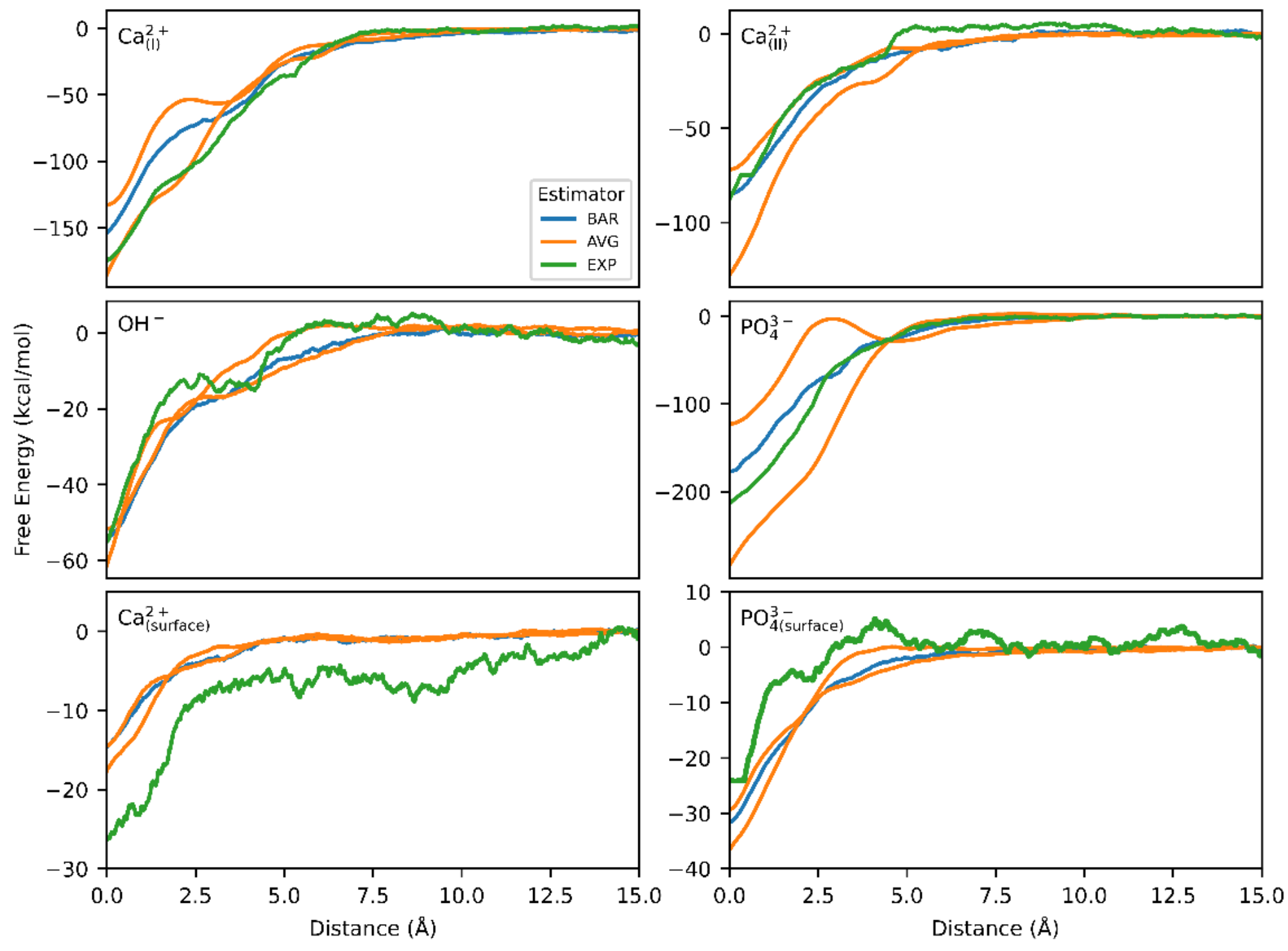

*Figure S 2 - Comparison of different free energy estimators with the reversible work for pulling different ions from the (001) facet of HAp at pH7.*

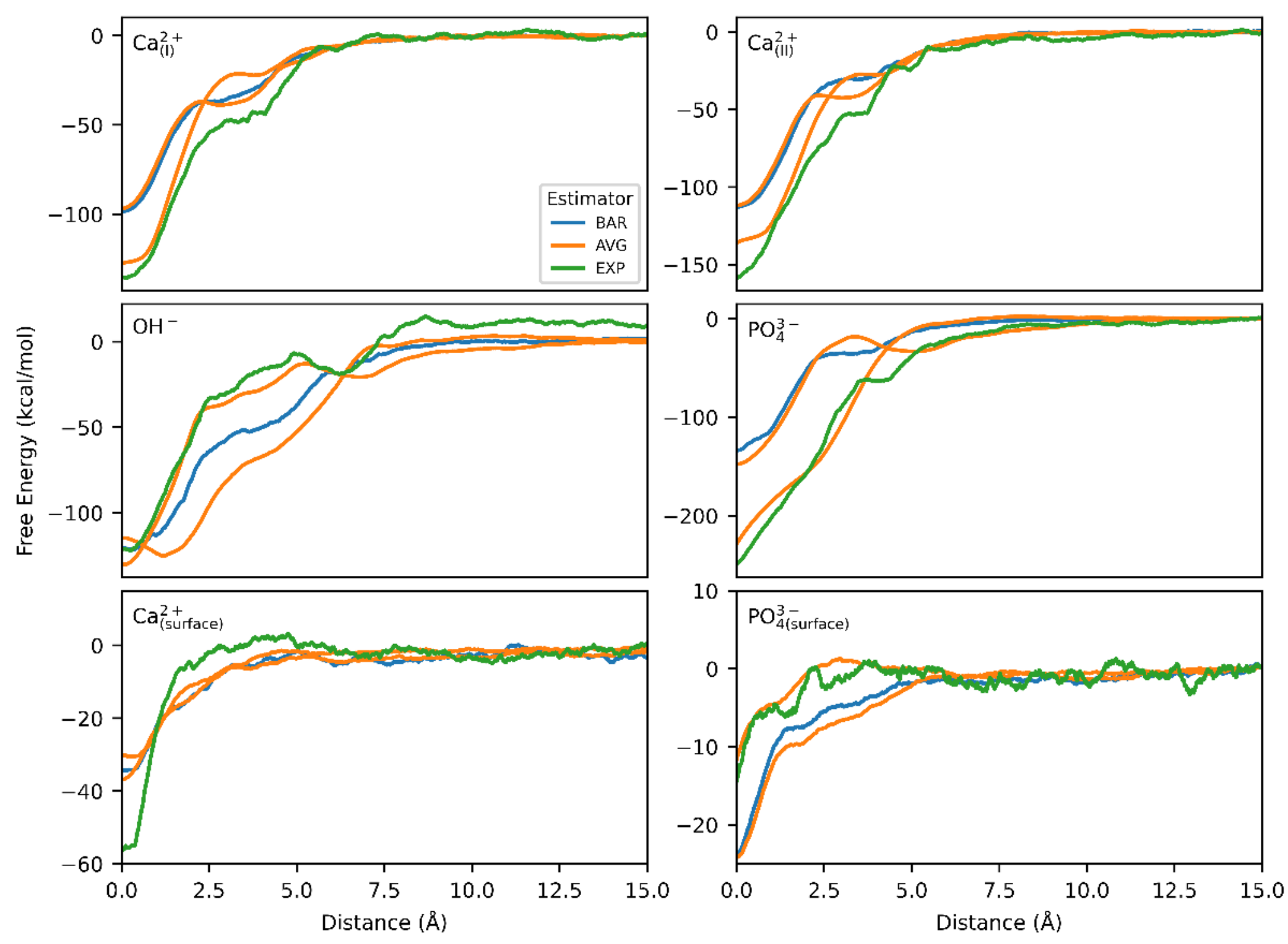

*Figure S 3 - Comparison of different free energy estimators with the reversible work for pulling different ions from the (010) facet of HAp at pH7.*

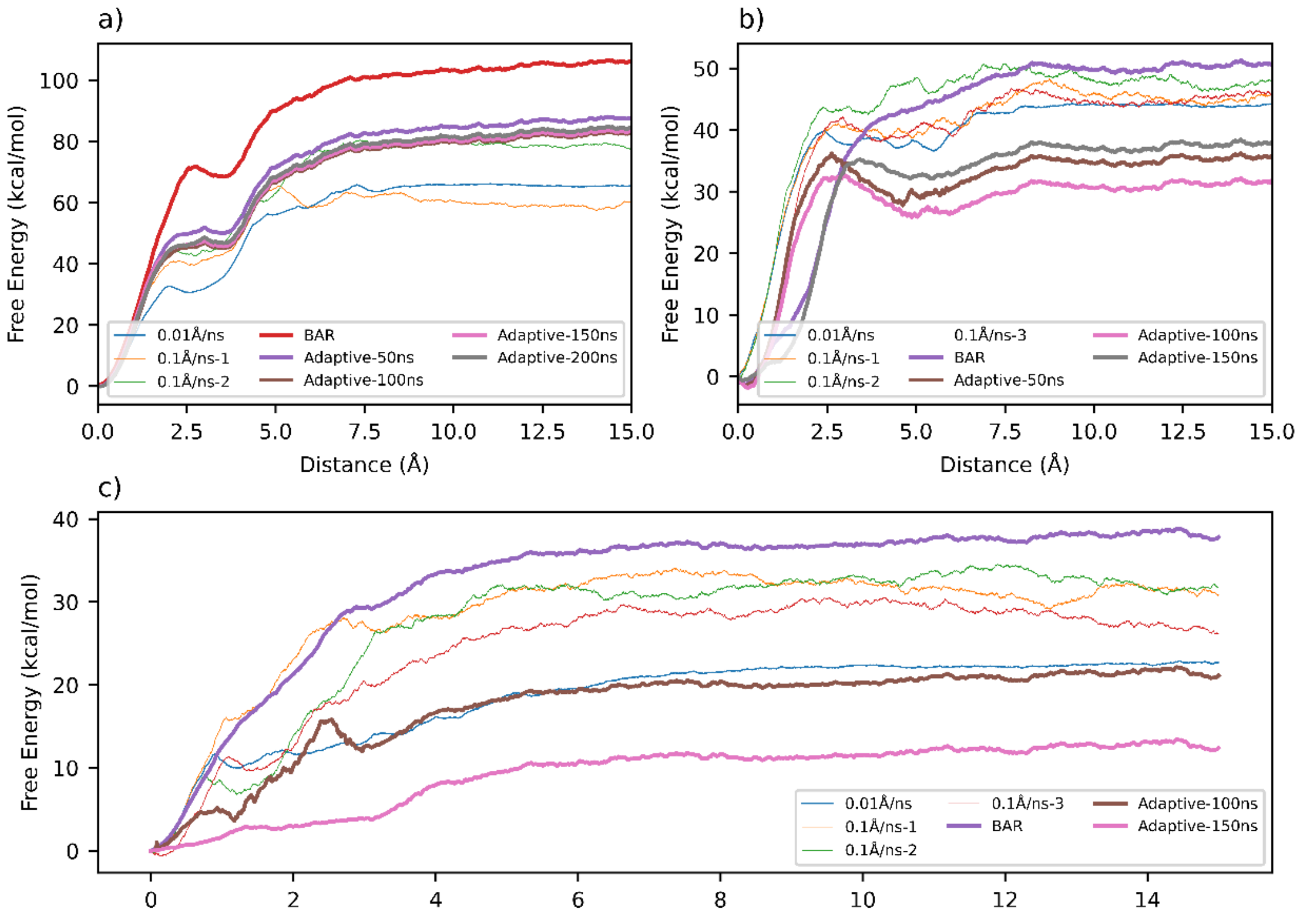

*Figure S 4 - Comparison of the adaptive-BAR simulation with the reversible pulling results. a) Pulling $Ca^{2+}$(I) from 010-pH5, b) $OH^-$ from 001 -pH5 and c) $PO_4^{3-}$(surface) from 001-pH7.*